\documentclass[aps,prl,superscriptaddress,preprint]{revtex4-1} % for PRL format
\usepackage{amsmath}
\usepackage{epstopdf}
\usepackage{graphicx}
\usepackage{bm,color,bbm}
\usepackage{hyperref, mathtools}
\usepackage{caption}
\usepackage{subcaption}

\newcommand{\beq}{\begin{equation}}
\newcommand{\eeq}{\end{equation}}
\newcommand{\bqa}{\begin{eqnarray}}
\newcommand{\eqa}{\end{eqnarray}}

\begin{document}

\preprint{}

\title{The Relationship Between Discrete and Continuous Entropy in EPR-Steering Inequalities}


\author{James Schneeloch}
\affiliation{Department of Physics and Astronomy, University of Rochester, Rochester, NY 14627}

%\author{Curtis J. Broadbent}
%\affiliation{Department of Physics and Astronomy, University of Rochester, Rochester, NY 14627}
%\affiliation{Rochester Theory Center, University of Rochester, Rochester, NY 14627}
%\author{John C.  Howell}
%\affiliation{Department of Physics and Astronomy, University of Rochester, Rochester, NY 14627}

%\email[]{jschneel@pas.rochester.edu, curtis@pas.rochester.edu}
%\thanks{}
%\affiliation{University of Rochester, Dept. of Physics and Astronomy}


\date{\today}

\begin{abstract}
This document expands upon the relationship between discrete and continuous entropy given in (Phys. Rev. Lett. \textbf{110} 130407), ``Violating Continuous Variable Einstein-Podolsky-Rosen Steering with Discrete Measurements". We provide a detailed derivation for the inequality relating the continuous conditional entropy to its discrete approximation, and show how this connection works between discrete and continuous entropic quantities in general. In addition, we use this connection to show how to derive the continuous variable Einstein-Podolsky-Rosen steering inequality with discrete measurements as seen in (Phys. Rev. Lett. \textbf{110} 130407), and make an additional comment which strengthens this result.
\end{abstract}

%\keywords{}

\maketitle


\section{Extended Proof of Entropy Connection Inequalities}
To derive our continuous variable Einstein-Podolsky-Rosen (EPR)-steering inequality \eqref{finalresult}\cite{Schneeloch2012}, we used a fundamental connection between continuous and discrete entropies \eqref{fundamentalconnection} to show that any two continuous random variables $x$ and $y$ that can be discretized into equally spaced windows of size $\Delta x$ and $\Delta y$ satisfy the following inequality;
\begin{equation} \label{myineq1}
\boxed{h(y|x) \leq H(Y|X) + \log(\Delta y),}
\end{equation} 
where the base of the logarithms here and throughout this paper are equal to the base in which one chooses to measure the entropy.

\subsection{The Fundamental Connection between Continuous and Discrete Entropies}
Consider an experiment to measure random variable $x$ which can take the value of any real number with probability density $\rho(x)$. The experiment is only capable of measuring $x$ to discrete windows $X_{\ell}$ of size $\Delta x$. The probability of measuring $x$ to be within window $X_{\ell}$ is
\begin{equation}\label{discprob}
P(X_{\ell}) = \int\limits_{\Delta x_{\ell}} dx\;  \rho(x),
\end{equation}
where the region of integration $\Delta x_{\ell}$ is the range of values of $x$ between $x_{\ell} - \frac{1}{2}\Delta x$ and  $x_{\ell} + \frac{1}{2}\Delta x$, and $x_{\ell}$ is the value of $x$ at the center of the window $X_{\ell}$. The Shannon entropy of this discrete probability distribution is given by
\begin{equation}
H(X) = - \sum_{\ell} P(X_{\ell}) \log(P(X_{\ell})),
\end{equation}
and the Shannon entropy of the continuous probability density function $\rho(x)$ is expressed as
\begin{equation}
h(x) = - \int dx\;\rho(x)\log(\rho(x)).
\end{equation}
We now define the probability density function $\rho_{\ell}(x)$ as the probability distribution of $x$ conditioned on having been measured within window $X_{\ell}$. The continuous entropy $h_{\ell}(x)$ is defined as the entropy of $\rho_{\ell}(x)$ where
\begin{equation}
\rho_{\ell}(x) = \frac{\rho(x)}{P(X_{\ell})}
\end{equation}
for all values of $x$ in the window $X_{\ell}$, and is zero otherwise.

By breaking up the continuous entropy $h(x)$ into a sum over all windows,
\begin{equation}
h(x) = - \sum_{\ell}\int\limits_{\Delta x_{\ell}} dx\;\rho(x)\log(\rho(x)),
\end{equation}
and expressing $h(x)$ in terms of $P(X_{\ell})$ and $\rho_{\ell}(x)$,
\begin{align}
h(x) &= - \sum_{\ell}\int\limits_{\Delta x_{\ell}} dx\;\big(\rho_{\ell}(x) P(X_{\ell})\big)\log\big(\rho_{\ell}(x) P(X_{\ell})\big)\nonumber\\
&=- \sum_{\ell}P(X_{\ell})\int\limits_{\Delta x_{\ell}} dx\;\rho_{\ell}(x) \bigg(\log(\rho_{\ell}(x)) + \log( P(X_{\ell})\bigg),
\end{align}
and then in terms of $h_{\ell}(x)$,
\begin{equation}
h(x)=- \sum_{\ell}P(X_{\ell})\int\limits_{\Delta x_{\ell}} dx\;\rho_{\ell}(x) \log(\rho_{\ell}(x)) \;-\; \sum_{\ell}P(X_{\ell})\log( P(X_{\ell})),
\end{equation}
we obtain the fundamental connection between discrete and continuous entropies;
\begin{equation}\label{fundamentalconnection}
\boxed{h(x) = \sum_{\ell} P(X_{\ell}) h_{\ell}(x) + H(X).}
\end{equation}

\subsection{Continuing the Extended Proof}
This connection \eqref{fundamentalconnection} exists for joint entropies as well as for marginal entropies. Using this, we now define $h_{\ell m}(x,y)$ as the entropy of the joint distribution $\rho_{\ell m}(x,y)$ in which $x$ is conditioned on being measured within window $X_{\ell}$ and $y$ is conditioned on being measured within window $Y_{m}$.

The conditional entropies $h(y|x)$ and $H(Y|X)$ are defined as differences between joint and marginal entropies \cite{Cover2006},
\begin{subequations}
\begin{align}\label{def}
h(y|x)&\equiv h(x,y) - h(x),\\
H(Y|X)&\equiv H(X,Y) - H(X).
\end{align}
\end{subequations}
By using the fundamental connection \eqref{fundamentalconnection} for both marginal and joint entropies, along with the definition of conditional entropy \eqref{def}, we can show that
\begin{align}\label{hygx}
h(y|x) &= \sum_{\ell,m} P(X_{\ell},Y_{m}) h_{\ell m}(x,y) - \sum_{\ell} P(X_{\ell}) h_{\ell}(x) + H(Y|X)\nonumber\\
&= \sum_{\ell} P(X_{\ell})\sum_{m}P(Y_{m}|X_{\ell}) h_{\ell m}(x,y) - \sum_{\ell} P(X_{\ell}) h_{\ell}(x) + H(Y|X).
\end{align}

Conditioning on additional events on average reduces the entropy. This is a consequence both of Jensen's inequality and the fact that the entropy is a concave function \cite{Cover2006}. As such, we can say that where
\begin{equation}
\rho(x) = \sum_{\ell} P(X_{\ell})\rho_{\ell}(x),
\end{equation}
we have the inequality
\begin{equation}
h(x) \geq \sum_{\ell} P(X_{\ell})h_{\ell}(x).
\end{equation}
Similarly, where \footnote{$\rho_{\ell m}(x)$ is just what you get when you integrate over all values of $y$ the probability density function $\rho_{\ell m}(x,y)$.}
\begin{equation}
\rho_{\ell}(x) = \sum_{m} P(Y_{m}|X_{\ell}) \rho_{\ell m}(x),
\end{equation}
we have the inequality
\begin{equation}\label{thisRelation}
h_{\ell}(x) \geq \sum_{m} P(Y_{m}|X_{\ell}) h_{\ell m}(x).
\end{equation}
When this relation \eqref{thisRelation} is substituted as a minimum value of $h_{\ell}(x)$ in the expression for $h(y|x)$, \eqref{hygx}, it can be shown that
\begin{equation}
h(y|x) \leq  \sum_{\ell,m} P(X_{\ell},Y_{m}) h_{\ell m}(y|x) + H(Y|X).
\end{equation}
The uniform distribution maximizes the entropy, so that when all windows $\Delta y_{m}$ are of equal size, we have $h_{\ell m}(y|x)\leq \log(\Delta y)$, which completes our proof of \eqref{myineq1}.

This approach allows us to summarize the connection between continuous and discrete entropic quantities including the mutual information, $h(x:y)$ defined as $h(x)+h(y)-h(x,y)$. In short:
\begin{align*}
h(x)\leq H(X) + \log(\Delta x) ,\\
h(x,y) \leq H(X,Y) + \log(\Delta x \Delta y), \\
h(x|y) \leq H(X|Y) + \log(\Delta x), \\
h(x:y) \geq H(X:Y) .
\end{align*}
This also translates to three or more variables easily, giving us
\begin{align*}
h(x,y,z)\leq H(X,Y,Z) + \log(\Delta x \Delta y \Delta z) ,\\
h(x,y|z)\leq H(X,Y|Z) + \log(\Delta x \Delta y), \\
h(x|y,z)\leq H(X|Y,Z) + \log(\Delta x), \\
h(x:y,z)\geq H(X:Y,Z) ,
\end{align*}
and in particular, we have
\begin{equation}\label{MondoResultLikeDude}
h(\vec{x}|\vec{y})\leq H(\vec{X}|\vec{Y}) +\log\bigg( \prod_{i=1}^{n}\Delta x_{i}\bigg),
\end{equation}
where $\vec{x}$ is a vector of the random variables $(x_{1},x_{2},...,x_{n})$.

It remains to be shown whether such a simple relationship exists between  the conditional mutual informations $h(x:y|z)$ and $H(X:Y|Z)$. Since the conditional mutual information $h(x:y|z)$ can be either greater or less than the ordinary mutual information, $h(x:y)$, \cite{Cover2006} the relationship between $h(x:y|z)$ and $H(X:Y|Z)$ may not be straightforward.

\section{Incorporating the Entropy Connection into our Steering Inequality}
Where $x_{Ai}$ and $x_{Bi}$ are another ordinary pair of random variables, we can substitute the larger discrete approximations \eqref{myineq1} for each continuous conditional entropy in the steering inequality derived by Walborn \emph{et.~al}~\cite{Walborn2011},
\begin{equation}
h(x_{Bi}|x_{Ai}) + h(k_{Bi}|k_{Ai}) \geq \log(\pi e),
\end{equation}
to derive our entropic EPR steering inequality suitable for experimental investigations of continuous variable entanglement. For a particular spatial degree of freedom $i\in\{1,...,n\}$, $($i.e. a particular dimension in space$)$ we've shown that
\begin{equation}\label{discWSE2}
H(X_{Bi}|X_{Ai}) + H(K_{Bi}|K_{Ai}) \geq \log \bigg(\frac{\pi e}{\Delta x_{Bi} \Delta k_{Bi}} \bigg).
\end{equation}
When different spatial degrees of freedom are statistically independent of one another, the entropies add, giving us the $n$-dimensional discrete steering inequality
\begin{equation}\label{finalresult}
\boxed{H(\vec{X}_{B}|\vec{X}_{A}) + H(\vec{K}_{B}|\vec{K}_{A}) \geq \sum_{i=1}^{n} \log \bigg(\frac{\pi e}{\Delta x_{Bi} \Delta k_{Bi}} \bigg).}
\end{equation}

\subsection{Comment on Steering Inequality Derivation}
Though the steering inequality just arrived at \eqref{finalresult} is quite valid, it is in fact stronger than the prior derivation suggests. The prior derivation requires the assumption that different spatial degrees of freedom be independent of one another, so that the steering inequality in multiple dimensions \eqref{finalresult} is just the sum of the steering inequalities in each dimension \eqref{discWSE2}. We can in fact do away with this assumption, by noting that without any additional assumptions, Walborn \emph{et.~al's} steering inequality \cite{Walborn2011} in $n$ spatial degrees of freedom is
\begin{equation}\label{walbineqnd}
h(\vec{x}^{B}|\vec{x}^{A}) + h(\vec{k}^{B}|\vec{k}^{A}) \geq n \log(\pi e),
\end{equation}
because Bia{\l}ynicki-Birula and Mycielski's entropic uncertainty relation \cite{BiałynickiBirula1975} in $n$ spatial degrees of freedom is
\begin{equation}
h(\vec{x}) + h(\vec{k})\geq n \log(\pi e),
\end{equation}
which requires no additional assumptions to prove.

Knowing this, we can use the steering inequality \eqref{walbineqnd} and the connection \eqref{MondoResultLikeDude} to prove our steering inequality \eqref{finalresult} without any extra assumptions. This strengthens the significance of our experimental violation of \eqref{finalresult}, so that, loopholes aside, it is a solid demonstration of EPR-steering without special caveats.

\bibliography{EPRbib10}

\begin{thebibliography}{5}%
\makeatletter
\providecommand \@ifxundefined [1]{%
 \@ifx{#1\undefined}
}%
\providecommand \@ifnum [1]{%
 \ifnum #1\expandafter \@firstoftwo
 \else \expandafter \@secondoftwo
 \fi
}%
\providecommand \@ifx [1]{%
 \ifx #1\expandafter \@firstoftwo
 \else \expandafter \@secondoftwo
 \fi
}%
\providecommand \natexlab [1]{#1}%
\providecommand \enquote  [1]{``#1''}%
\providecommand \bibnamefont  [1]{#1}%
\providecommand \bibfnamefont [1]{#1}%
\providecommand \citenamefont [1]{#1}%
\providecommand \href@noop [0]{\@secondoftwo}%
\providecommand \href [0]{\begingroup \@sanitize@url \@href}%
\providecommand \@href[1]{\@@startlink{#1}\@@href}%
\providecommand \@@href[1]{\endgroup#1\@@endlink}%
\providecommand \@sanitize@url [0]{\catcode `\\12\catcode `\$12\catcode
  `\&12\catcode `\#12\catcode `\^12\catcode `\_12\catcode `\%12\relax}%
\providecommand \@@startlink[1]{}%
\providecommand \@@endlink[0]{}%
\providecommand \url  [0]{\begingroup\@sanitize@url \@url }%
\providecommand \@url [1]{\endgroup\@href {#1}{\urlprefix }}%
\providecommand \urlprefix  [0]{URL }%
\providecommand \Eprint [0]{\href }%
\providecommand \doibase [0]{http://dx.doi.org/}%
\providecommand \selectlanguage [0]{\@gobble}%
\providecommand \bibinfo  [0]{\@secondoftwo}%
\providecommand \bibfield  [0]{\@secondoftwo}%
\providecommand \translation [1]{[#1]}%
\providecommand \BibitemOpen [0]{}%
\providecommand \bibitemStop [0]{}%
\providecommand \bibitemNoStop [0]{.\EOS\space}%
\providecommand \EOS [0]{\spacefactor3000\relax}%
\providecommand \BibitemShut  [1]{\csname bibitem#1\endcsname}%
\let\auto@bib@innerbib\@empty
%</preamble>
\bibitem [{\citenamefont {Schneeloch}\ \emph {et~al.}(2013)\citenamefont
  {Schneeloch}, \citenamefont {Dixon}, \citenamefont {Howland}, \citenamefont
  {Broadbent},\ and\ \citenamefont {Howell}}]{Schneeloch2012}%
  \BibitemOpen
  \bibfield  {author} {\bibinfo {author} {\bibfnamefont {J.}~\bibnamefont
  {Schneeloch}}, \bibinfo {author} {\bibfnamefont {P.~B.}\ \bibnamefont
  {Dixon}}, \bibinfo {author} {\bibfnamefont {G.~A.}\ \bibnamefont {Howland}},
  \bibinfo {author} {\bibfnamefont {C.~J.}\ \bibnamefont {Broadbent}}, \ and\
  \bibinfo {author} {\bibfnamefont {J.~C.}\ \bibnamefont {Howell}},\ }\href
  {\doibase 10.1103/PhysRevLett.110.130407} {\bibfield  {journal} {\bibinfo
  {journal} {Phys. Rev. Lett.}\ }\textbf {\bibinfo {volume} {110}},\ \bibinfo
  {pages} {130407} (\bibinfo {year} {2013})}\BibitemShut {NoStop}%
\bibitem [{\citenamefont {Cover}\ and\ \citenamefont
  {Thomas}(2006)}]{Cover2006}%
  \BibitemOpen
  \bibfield  {author} {\bibinfo {author} {\bibfnamefont {T.~M.}\ \bibnamefont
  {Cover}}\ and\ \bibinfo {author} {\bibfnamefont {J.~A.}\ \bibnamefont
  {Thomas}},\ }\href@noop {} {\emph {\bibinfo {title} {Elements of Information
  Theory}}},\ \bibinfo {edition} {2nd}\ ed.\ (\bibinfo  {publisher} {Wiley and
  Sons},\ \bibinfo {address} {New York},\ \bibinfo {year} {2006})\BibitemShut
  {NoStop}%
\bibitem [{Note1()}]{Note1}%
  \BibitemOpen
  \bibinfo {note} {$\rho _{\ell m}(x)$ is just what you get when you integrate
  over all values of $y$ the probability density function $\rho _{\ell
  m}(x,y)$.}\BibitemShut {Stop}%
\bibitem [{\citenamefont {Walborn}\ \emph {et~al.}(2011)\citenamefont
  {Walborn}, \citenamefont {Salles}, \citenamefont {Gomes}, \citenamefont
  {Toscano},\ and\ \citenamefont {Souto~Ribeiro}}]{Walborn2011}%
  \BibitemOpen
  \bibfield  {author} {\bibinfo {author} {\bibfnamefont {S.~P.}\ \bibnamefont
  {Walborn}}, \bibinfo {author} {\bibfnamefont {A.}~\bibnamefont {Salles}},
  \bibinfo {author} {\bibfnamefont {R.~M.}\ \bibnamefont {Gomes}}, \bibinfo
  {author} {\bibfnamefont {F.}~\bibnamefont {Toscano}}, \ and\ \bibinfo
  {author} {\bibfnamefont {P.~H.}\ \bibnamefont {Souto~Ribeiro}},\ }\href
  {\doibase 10.1103/PhysRevLett.106.130402} {\bibfield  {journal} {\bibinfo
  {journal} {Phys. Rev. Lett.}\ }\textbf {\bibinfo {volume} {106}},\ \bibinfo
  {pages} {130402} (\bibinfo {year} {2011})}\BibitemShut {NoStop}%
\bibitem [{\citenamefont {Bia{\l}ynicki-Birula}\ and\ \citenamefont
  {Mycielski}(1975)}]{BiałynickiBirula1975}%
  \BibitemOpen
  \bibfield  {author} {\bibinfo {author} {\bibfnamefont {I.}~\bibnamefont
  {Bia{\l}ynicki-Birula}}\ and\ \bibinfo {author} {\bibfnamefont
  {J.}~\bibnamefont {Mycielski}},\ }\href
  {http://dx.doi.org/10.1007/BF01608825} {\bibfield  {journal} {\bibinfo
  {journal} {Communications in Mathematical Physics}\ }\textbf {\bibinfo
  {volume} {44}},\ \bibinfo {pages} {129} (\bibinfo {year} {1975})},\ \bibinfo
  {note} {10.1007/BF01608825}\BibitemShut {NoStop}%
\end{thebibliography}%

\end{document}